\documentclass[10pt,conference]{IEEEtran}
\usepackage{amssymb}
\usepackage{amsmath}
\usepackage[dvips]{graphicx}
\usepackage{bm}

\title{Unusual Kondo physics in a Co impurity atom embedded in noble-metal chains}
\author{\IEEEauthorblockN{ S. Di Napoli\IEEEauthorrefmark{1}, M.A. Barral\IEEEauthorrefmark{1}, P. Roura-Bas\IEEEauthorrefmark{1},
   A.A. Aligia\IEEEauthorrefmark{2}, Y. Mokrousov\IEEEauthorrefmark{3} and A.M. Llois\IEEEauthorrefmark{1}}
\IEEEauthorblockA{\IEEEauthorrefmark{1}Departamento de F\'{\i}sica de la Materia Condensada, CAC-CNEA\\
 Avenida General Paz 1499, (1650) San Mart\'{\i}n, Pcia. de Buenos Aires, Argentina and \\
Consejo Nacional de Investigaciones Cient\'{\i}ficas y T\'ecnicas, CONICET, Buenos Aires, Argentina}
\IEEEauthorblockA{\IEEEauthorrefmark{2} Centro At\'omico Bariloche and Instituto Balseiro, Comisi\'on Nacional de
Energ\'{\i}a At\'omica, 8400 Bariloche}
\IEEEauthorblockA{\IEEEauthorrefmark{3}Peter Gr\"unberg Institut and Institute for Advanced Simulation,
Forschungszentrum J\"ulich and JARA, 52425 J\"ulich, Germany}}

\begin{document}
\maketitle

\begin{abstract}
We analyze the conduction bands of the one dimensional noble-metal chains 
that contain a Co magnetic impurity  
by means of \textit{ab initio} calculations. We compare the results 
obtained for Cu and Ag pure chains, as well as O doped Cu, Ag and Au chains 
with those previously found for Au pure chains.
We find similar results in the case of Cu and Au hosts, whereas for Ag chains a 
different behavior is obtained.  Differences and similarities among the 
different systems are analyzed by comparing the electronic structure of the 
three noble-metal hosts. 
The \textit{d} orbitals of Cu chains at the Fermi level have the same symmetry 
as in the case of Au chains. 
These orbitals hybridize with the corresponding ones of the Co impurity, 
giving rise to the possibility of
exhibiting a two-channel Kondo physics. 
\end{abstract}

\section{Introduction}
One of the most exciting  areas of research in the last few years, is the 
realization
of pure one dimensional (1D) systems, as they open the possibility of 
studying not 
only theoretically but, also experimentally, the characteristic features
triggered by low-dimension,
that involved  electronic transport, correlation and magnetic properties. 
Atomic-size 
contacts and nanowires can nowadays be obtained by scanning tunneling 
microscope or in mechanically controllable break junctions experiments 
(MCBJ)~\cite{Ruitenbeek01}. It is with 
this last technique that it has been possible to create monoatomic 
chains of Ir~\cite{Ryu06},
 Pt~\cite{Ruitenbeek01} and Au~\cite{Ohnishi98,Ruitenbeek98,Rubio01}. 
Recently, it has
been also demonstrated that impurity-assisted chain growth leads to a 
strongly enhanced tendency towards chain formation, when compared to pure 
noble-metal (NM) chains. In particular, 
$p$-like impurities, like N or O, lead to high chain elongation probabilities 
due to the strong directional 
bondings~\cite{Bahn02,Cakir11,Novaes06,Thijssen06,Alex08,Dinapoli12}.
Given these experimental outstanding achievements, a new generation of 
nanodevices can in principle be 
conceived and designed to be used as tools for detecting nanomagnetism 
by conductance measurements through atomic metal contacts. One can 
indirectly sense the presence of magnetism by detecting zero-bias
anomalies, while doing conductance measurements through an atomic contact. 
The origin of these zero-bias anomalies is usually due to Kondo screening of the spin, 
if a magnetic impurity is 
bridging the contact. 

One of the prototype magnetic impurities in idealized models and 
computational simulations are Co atoms,
as they develop a wide range of spin moments depending on the metal host
and the dimensionality
with usually high enough Kondo temperatures. 
Co impurities in Au, Ag and Cu
surfaces have been extensively studied for several years and they are still 
at the focus of research.
Recently, H. Pr\"user and co-workers analyzed the Kondo resonance at Co 
atoms buried below a Cu(100) surface by mapping 
the local density of states of the surface itinerant electrons~\cite{Pruser12} 
and B. Surer \textit{et al.} developed a multiorbital Kondo model to 
investigate the physics of Co atoms in Cu
hosts~\cite{Surer12}. On the other hand, Co impurities in noble-metal 
chains are interesting in order to study how the reduced dimensionality 
could lead to different and novel 
transport and magnetic properties and in this way eventually form part of a 
nanomagnetism detecting nanodevice.
 
In a previous work, we found that 
non-Fermi liquid behavior qualitatively similar to that of
the two-channel Kondo (2CK) model is expected in a system of a Co impurity 
embedded in a Au chain which is in contact with Au leads, that break strongly enough 
the symmetry of the chain reducing it to a four-fold one~\cite{Dinapoli13}.
The proposed setup might be realized in break junction experiments if the symmetry of 
the contacts can be controlled or in STM experiments on (100) noble-metal surfaces.

In this contribution, we want to analyze if the same physical trends, that
is, the presence of the 2CK physics and the non Fermi liquid behavior, 
could be obtained when instead of Au chains we introduce a Co
impurity in Cu and Ag chains. As mentioned previously, these last noble-metal  
chains could be
stabilized by introducing light environmental non-magnetic impurities in 
the experimental atmosphere~\cite{Dinapoli12}. With this purpose we obtain and
analyze the electronic structure of these metallic systems. In the case of 
Cu chains we check if the presence of
stabilizing O atoms could destroy or not the eventual presence of a 2CK effect.

The paper is organized as follows. In section \ref{kondo} we briefly recall 
the ground state properties of a Co impurity in
a Au host, related to the Kondo effect. In section \ref{results}  we provide 
the details of our DFT first-principles calculations and 
analyze the bandstructure of the different noble-metal hosts and the
consequences of these structures on the systems with an embedded Co impurity. 
Finally, a summary and conclusions are given in section \ref{conclusions}. 
 
\section{Co impurity within a Au chain: Kondo effect}
\label{kondo}

A magnetic impurity of Co embedded in a Au chain, 
under a four-fold symmetry breaking field $B$
exhibits different
transport properties depending on the specific geometry of the 
leads~\cite{Dinapoli13}. The Co atom embedded in a Au chain has a total spin 
$S=3/2$ in a $3d^7$ configuration. One $d$ hole is shared by the half filled 
$3d_{xy}$ and $3d_{x^2-y^2}$ ($\Delta_4$-symmetry),
while the other two are in the empty and degenerate $3d_{xz}$ and $3d_{yz}$ 
($\Delta_3$-symmetry). On the other hand, the $5d_{xz}$ and 
$5d_{yz}$ of the pure Au chains are also degenerate and close to the 
Fermi level and will be pushed up by the presence of O impurities. 
Therefore, these represent two identical conduction bands 
that can screen the $S=1/2$ spin of the
$3d_{xz}$ and $3d_{yz}$ holes of Co. On the contrary, there is no density 
of states of Au with $\Delta_4$ symmetry at
the Fermi level that could be hybridized with
states of the same symmetry at the Co site, 
leading to a frozen charge in the $3d_{x^2-y^2}$ and $3d_{xy}$ levels.

In presence of $B$ and in absence of spin-orbit coupling, 
the microscopic model that describes the system consists of a spin 3/2
hybridized with two triplets of the $d_8$ configuration through two
conduction channels (the $5d_{xz}$ and $5d_{yz}$ of Au) \cite{Dinapoli13}.
The corresponding physics is similar to that of the underscreened 
Kondo or Anderson models~\cite{Aligia86,Mehta05}.
However, the spin-orbit coupling induces a splitting $D$ between the states 
with projection $M =\pm 3/2$ and $M =\pm 1/2$. This splitting has been calculated 
solving exactly the 3d$^7$ configuration of Co including all correlations of the 
d shell~\cite{Aligia10}. 
For a negative $D$, larger than the characteristic Kondo temperature, the doublet $M =\pm 3/2$ is
lower in energy than the $M =\pm 1/2$ one and, therefore, the two conduction bands can not change the local $M = -3/2$ into the $M = +3/2$. Thus, the 
spin flip process of the Kondo effect is inhibited. 

On the contrary, for positive $D$, which is the case for large enough $B$~\cite{Dinapoli13}, the $M =\pm 1/2$ doublet is below in energy
than the $M =\pm 3/2$ one, being overcompensated by the two $\Delta_3$ conduction leads.
This overcompensation is usually the appropriate scenario to obtain experimentally the well known 2CK physics~\cite{Andrei84,
Zarand06,Mitchell12,Oreg03}, and a similar physics is in fact confirmed by detailed calculations using the numerical renormalization group in the model \cite{Dinapoli13}. In particular the low-temperature entropy is ln(2)/2 and
the conductance through the device displays a $T^{1/2}$ behavior. 
However, the model presents some differences with the 2CK one. For example, the conductance at zero temperature is less than expected due to 
some degree of intermediate valence~\cite{Dinapoli13}.
 
While the properties of the ground state,
underscreened or overscreened Kondo physics, are determined by the symmetry properties of the leads and the strength of the four-fold crystal field $B$, 
the crucial point for 
observing the Kondo effect is the presence of $5d_{xz}$ and $5d_{yz}$ density of states of Au at the Fermi level. It is not
obvious that the same trends can be obtained when chains of other materials are used.

\section{Results}
\label{results}

We perform \textit{ab initio} calculations based on spin polarized density functional 
theory (SP-DFT) using the full potential linearized
augmented plane waves method, as implemented in the WIEN2K code~\cite{Wien}. 
The generalized gradient approximation for
the exchange and correlation potential in the parametrization of PBE and 
the augmented plane
waves local orbital basis are used. The cutoff parameter which gives the 
number of plane waves in the interstitial region 
is taken as Rmt*Kmax = 7, where Kmax is the value of the largest reciprocal 
lattice vector used in the plane waves’ expansion
and Rmt is the smallest muffin tin radius used. The number of \textbf{k} 
points in the Brillouin zone is enough in each case 
to obtain the desired energy  and charge precisions, namely 10$^{-4}$ Ry and 
10$^{-4}$e, respectively. 
The muffin-tin radii were set to $2.09$ bohr for Co, Cu, Ag and Au atoms 
and $1.52$ bohr for the O impurities.  

\begin{figure}
\includegraphics[width=1.0\columnwidth]{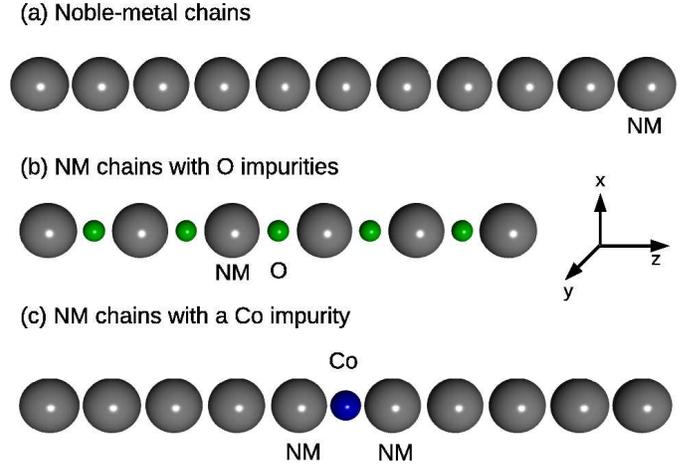}
\caption{(Color online) Schematic representation of the systems studied. The noble-metal (NM) atoms 
are represented by large grey spheres, the smaller green spheres represent the O atoms, while the blue sphere stands for 
the Co magnetic impurity.}
\label{setup}
\end{figure}

We set the coordinate system such that the chain axis is aligned along 
the $z$ axis, as schematically shown in Fig.~\ref{setup}.
For the noble-metal chains we 
consider one-atom unit cells (Fig.~\ref{setup}(a)) in a hexagonal lattice with $a=b=10$ bohr 
and $c=d_{NM-NM}^ {eq}$, where $d_{NM-NM}^ {eq}$ is the chain's equilibrium 
lattice constant in each case ($d_{NM-NM}^ {eq} = 4.40, 5.09$ and $4.93$ bohr for Cu, Ag and Au, respectively).
In O doped NM chains, we consider, for simplicity, that the presence of O atoms results in ...NM-O-NM-O... linear chain structures, to make clear the doping effect on the noble-metal d-bands.  The distances between the noble-metal to O atoms were relaxed in the $z$-direction, in all the cases (Fig.~\ref{setup}(b)).
Finally, for the NM chains plus the Co embedded impurity we consider an eleven 
atoms unit cell, (Fig.~\ref{setup}(c)), with the noble-metal to noble-metal distances, $d_{NM-NM}$,
set equal to the equilibrium chain's lattice constant 
($d_{NM-NM}=d_{NM-NM}^ {eq}$), while the 
noble-metal to Co impurity distance, $d_{NM-Co}$, is set equal to 
$(d_{NM-NM}^{eq}+d_{Co-Co}^{eq})/2$, where $d_{Co-Co}$ is the 
optimal Co-Co distance in a Co chain.

To check if Ag and Cu chains could give raise to similar physics as
Au chains in the presence of the Co impurities, we compare the band 
structure of Cu and Ag linear chains with that obtained for the Au one. 
The PDOS as well as the
bandstructure of the different
NM chains are presented in Fig.~2, where it is seen that
the $d$ bands of the Ag chain are
well below the Fermi level (second panel of Fig. 2), while the
Cu case is similar to the Au one,
showing $d_{xz,yz}$ ($\Delta_3$-symmetry) states
at the Fermi level.
For the three chains, the $\Delta_4$ orbitals are more localized than
the $\Delta_3$ ones, and the more extended
orbitals are those of $d_z^2$ ($\Delta_1$)-symmetry. 

From the spin polarized calculation of the electronic structure of a Co atom embedded
in the Cu and Ag chains, it is obtained that similarly to what  
happens with a Co impurity in the Au chain, the Co
atom presents a total spin $S = 3/2$. The Co atom exhibits three
spin down holes, two with $\Delta_3$ symmetry and the last one coming from
the half-filled degenerate $\Delta_4$ orbitals. 
Due to the fact that
the Co impurity has the same structure when embedded in the three studied
hosts, we can return now to the differences and similarities in the 
bandstructure of the different NM chains already discussed above.
As a direct result of the electronic structure of  Cu chains, we can 
say that they would present
the same Kondo behavior as the one expected for the corresponding Au system. 
The characteristic Kondo scale will, of course,
depend on the specific parameters of each system. On the contrary, the Ag host does not provide the appropriate environment 
for the screening of the angular momentum characteristic in the Kondo effect.
 
\begin{figure}
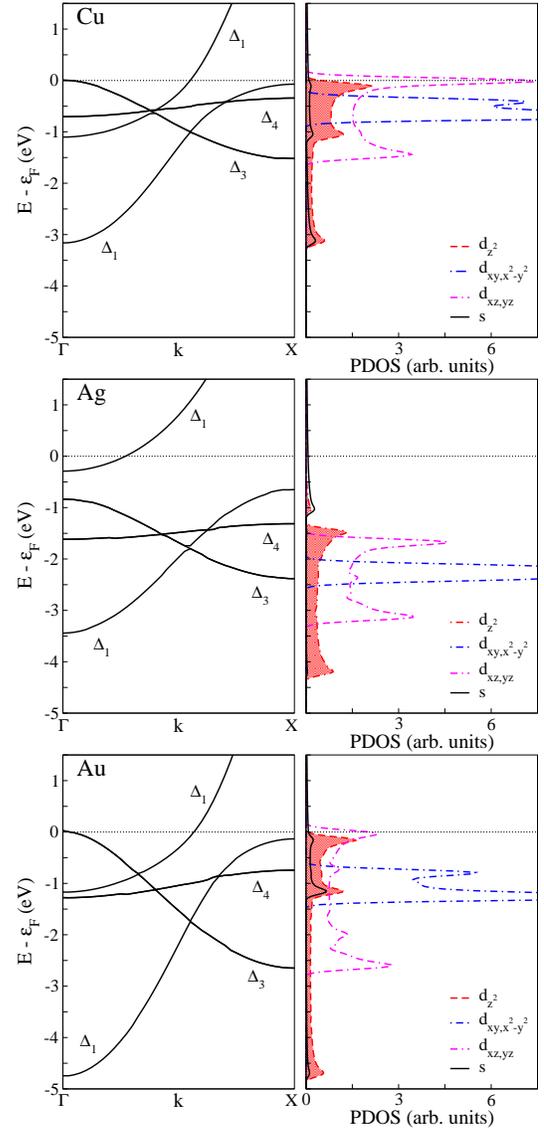

\begin{center}
\includegraphics[height=5cm]{Fig2-1.eps}
\includegraphics[height=5cm]{Fig2-2.eps}
\includegraphics[height=5cm]{Fig2-3.eps}
\caption{(Color online) Band structure and densities of states of Cu (top), 
Ag (middle) and Au (bottom) chains. The densities
of states are additionally decomposed into the contributions coming from 
$s$ and $d$ states of different symmetries.}
\end{center}
\label{densities}
\end{figure}

As mentioned in the introduction, impurity-assisted chain growth helps
to enhance the tendency towards noble-metal chain formation. 
The question arises then, what is the effect of O doping
on the Kondo trends of the NM chains with the embedded Co impurity?
After calculating the electronic structure of NM chains  doped with 
stabilizing oxygen atoms, we find that Co is again in a $S = 3/2$ spin
state, showing the same holes' symmetry as before. In the NM chains doped
with oxygen, we find that the $\Delta_3$ orbitals are pushed up towards the 
Fermi level due to hybridization of these states with the $p_{x,y}$ 
degenerate orbitals 
of the O atoms. This is in agreement with the
results presented in Ref.(~\cite{Dinapoli12}), where the \textit{ab initio} 
calculations were performed using a different code.
In Fig.~3 we show the densities of states for the three 
studied NM chains when doped with one O atom per NM atom. 
It is noteworthy to say that this is considered as a strong doping limit case just to strength the
role of the O atoms in the behavior of the NM d-bands.   Due to the metallic character of the 
chains as a consequence of the presence of partially filled s bands (which screen the on-site Coulomb repulsion),
we expect that the role of correlations is 
less important than for example in CuO linear chains in CuGeO$_3$, SrCuO$_2$~\cite{vekua} or in 
superconducting cuprates~\cite{Garces} and have a small impact on the relative position of the 
different bands.
In the high doping limit we observe that all the systems
present the $\Delta_3$ band crossing the Fermi level, giving rise to the 
possibility of hybridization of these states with
the $3d_{xz,yz}$ Co orbitals. In particular, it is clear from the second 
panel of Fig.~3 that the Ag doped chain
could now provide the proper environment for the development of the 
Kondo physics. Although this seems to be the main result of this
contribution, we also notice that the strength of the chain when doping 
with O impurities is even larger than the corresponding
one for the Au pure chains. (See Figure 7 of Ref.~\cite{Dinapoli12}). 
Therefore, the presence of O impurities plays an important role
in the case of Ag, it pushes up 
the $\Delta_3$ orbitals towards the Fermi level increasing the tendency
towards Kondo physics, and it strengthens the bonds within the Ag chain
enhancing its feasibility. The same conclusion holds for the Au and Cu 
doped chains.

For the sake of the comparison with the previous study in Au chains, in the case of the Cu pure chain 
we also added a four-fold breaking 
symmetry FCC lead. We obtain that the splitting between the $3d_{x^2-y^2}$
and the $3d_{xy}$ orbitals is large enough to produce a positive value 
of $D$, in agreement with the one obtained for the Au pure chains 
reported in Ref.~\cite{Dinapoli13}. Thus we expect that, in this particular 
case, the ground state will present the physics of the
overscreened Kondo model.

\begin{figure}
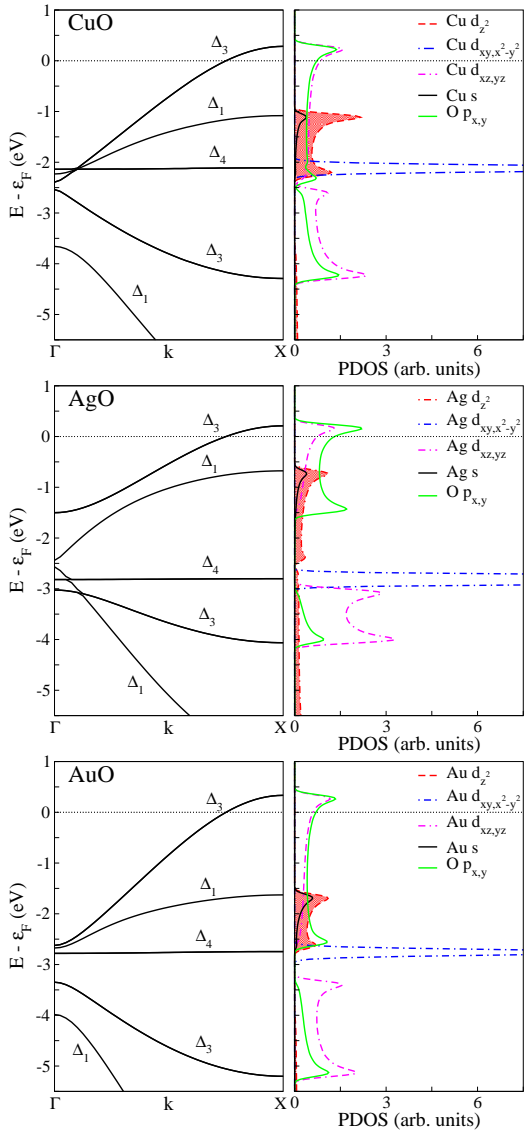

\begin{center}
\includegraphics[height=5cm]{Fig3-1.eps}
\includegraphics[height=5cm]{Fig3-2.eps}
\includegraphics[height=5cm]{Fig3-3.eps}
\caption{(Color online) Band structure and densities of states of NM chains doped with the O atoms. The densities
of states are additionally decomposed into the contributions coming from $s$ and $d$ states of different symmetries.}
\end{center}
\label{oxygen}
\end{figure}

\section{Conclusions}
\label{conclusions}

In this contribution we analyze the necessary conditions to obtain the Kondo physics in noble-metal chains with a Co embedded impurity. 
We find that Cu and Au leads behave electronically in a similar way in the presence of a Co impurity. The characteristic Kondo scale
will depend on the specific parameters of each system. On the contrary, Ag chains lack the necessary $\Delta_3$ symmetry at the Fermi
level which should provide the screening of the angular momentum, characteristic of the Kondo effect.
In an atmosphere of O atoms, the probability of developing the Kondo physics is enhanced in the Ag chains, while the Cu and Au chains increase
their mechanical stability and also improve the 2CK existence conditions. 
The physics of the overscreened Kondo model and non Fermi liquid properties are also foreseen in the case of Cu pure chains when adding a 
four-fold breaking symmetry field in the leads.

This work was partially supported by PIP 11220080101821, PIP 00258 of
CONICET, and PICT R1776 of the ANPCyT, Argentina.

\bibliographystyle{IEEEtran}
\bibliography{chains}

\end{document}